\documentclass[conference]{IEEEtran}
\IEEEoverridecommandlockouts

\usepackage{cite}
\usepackage{graphicx}
\usepackage{dcolumn}
\usepackage{bm}
\usepackage{amsmath, amssymb, amsthm}
\usepackage{mathtools}
\allowdisplaybreaks
\usepackage{physics}
\usepackage[caption=false]{subfig}
\usepackage{booktabs}
\usepackage{float}
\usepackage{siunitx}
\usepackage{comment}
\usepackage{url}
\usepackage{listings}
\usepackage[T1]{fontenc}
\usepackage{hyperref}
\usepackage{tikz}
\usetikzlibrary{arrows.meta,calc,fit,backgrounds,positioning}

\newcommand{\mc}[1]{\mathcal{#1}}

\begin{document}

\title{Entanglement Distillation and Swapping Scheduling in Quantum Repeaters with Noisy Memories}

\author{
\IEEEauthorblockN{
Siddharth Chander\IEEEauthorrefmark{1},
Xinan Chen\IEEEauthorrefmark{2},
and Allen Zang\IEEEauthorrefmark{3}
}

\IEEEauthorblockA{
\IEEEauthorrefmark{1}Department of Computer Science, California Institute of Technology, Pasadena, California 91125, USA\\
\IEEEauthorrefmark{2}Department of Electrical and Computer Engineering, University of Illinois at Urbana-Champaign, Urbana, Illinois 61801, USA\\
\IEEEauthorrefmark{3}Pritzker School of Molecular Engineering, University of Chicago, Chicago, Illinois 60637, USA\\
schander@caltech.edu, xchen146@illinois.edu, yzang@uchicago.edu
}
}

\maketitle

\begin{abstract}
    Entanglement distillation and entanglement swapping have been extensively studied assuming perfect quantum memories. However, near-term quantum networks will be fundamentally limited by quantum memories with a finite coherence time, resulting in complex choices for the timing and ordering of these operations. In this work, we study entanglement distillation and entanglement swapping at the level of the elementary building blocks of noisy quantum repeater networks, with the goal of elucidating the fundamental tradeoffs induced by memory decoherence. First, we focus on a minimal one-hop setting, where we analytically compare ``distill-as-soon-as-possible'' and ``distill-as-late-as-possible'' strategies against a baseline strategy that simply discards the older entangled state. We find that in the low memory coherence time regime, discarding the older entangled state achieves higher expected output fidelity, while in the high coherence time regime, delaying distillation until the end achieves the highest expected output fidelity and, over most of the deadline range, the highest weighted coherent information, at the expense of a lower success probability. We then extend our analysis to two-hop repeater chains using Monte Carlo simulation. In this setting, we find that the highest weighted coherent information is achieved by strategies that defer distillation to the end of the time window, with \textsf{Distill-ALAP-then-Swap-ALAP} and \textsf{Swap-ASAP-then-Distill-ALAP} leading at different operating points, while \textsf{Discard-Oldest-then-Swap} never reaches positive weighted coherent information in any regime tested. Together, these results clarify how decoherence reshapes the optimal operation timing in quantum networks and provide instructive insights into the link-level principles that govern larger-scale architectures.
\end{abstract}

\section{Introduction}
Establishing quantum entanglement between remote nodes is an essential ingredient for the realization of quantum networks~\cite{kimble2008quantum,wehner2018quantum}. Quantum repeaters~\cite{azuma2023quantum} represent a critical technological foundation for extending entanglement across long distances where photon loss and decoherence otherwise render direct transmission impractical. Each repeater node serves as both a quantum memory array and a site for entanglement distillation and swapping. In the near term, quantum systems will suffer not only from imperfect operations~\cite{briegel1998quantum,dur1999quantum} but also from limited entanglement generation rates and quantum memories coherence times. As such, there has been strong interest in optimizing the performance of quantum repeaters with these constraints in mind. 

The use of a memory cutoff time has been extensively studied in the literature to overcome memory noise in realistic quantum repeaters. Any entangled state stored longer than this threshold is declared useless and discarded. Within this line of work, Ref.~\cite{santra2019quantum} proposed the optimized buffer time protocol (OBP), which enforces a memory cutoff time that optimizes the entanglement of formation of the expected final end-to-end entangled state, showing that the entanglement of formation can be improved by several orders of magnitude compared to canonical protocols that do not optimize the buffer time. Building on this, Refs.~\cite{khatri2021policies} and~\cite{inesta2023optimal} brought Markov decision process methods to bear on the problem, characterizing optimal cutoff strategies at the elementary link level and finding policies that minimize expected entanglement delivery time across a repeater chain, respectively.

However, memory cutoff alone is insufficient when the initial fidelities of raw entangled states are low or when target fidelities are high, and in such regimes entanglement distillation is essential. Despite growing interest, the relationship between distillation, swapping, and memory decoherence remains incompletely understood. Ref.~\cite{zang2023entanglement} extends the OBP by including entanglement distillation performed as soon as possible with imperfect operations. Ref.~\cite{zang2025entanglement} analytically investigates the optimal time to perform entanglement distillation in noisy memories, while Ref.~\cite{zang2025no} proves a fundamental limitation on entanglement distillation in noisy memories. Haldar \textit{et al.}~\cite{haldar2025reducing} study multiplexed quantum repeaters with both swapping and distillation capabilities, demonstrating that quasi-local policies can reduce classical communication costs considerably. However, an important question remains unaddressed: Given that both distillation and swapping must be performed under memory constraints, in what order and at what time should each operation be executed? This is the central question motivating the present work. 
 
In this work, we provide an analysis of how memory decoherence shapes the optimal timing and ordering of distillation and swapping in multiplexed quantum repeaters. In order to isolate the main tradeoffs caused by memory decoherence, we focus on the elementary building blocks of quantum repeater networks: a one-hop link and a two-hop repeater chain. These building blocks can be used to construct longer chains hierarchically~\cite{briegel1998quantum,santra2019quantum,zang2023entanglement}, so the operational principles established here carry over to larger-scale architectures.
 
We assume the repeater chain is multiplexed (see Fig.~\ref{fig:diagrams}). Heralded entanglement generation between two pairs of noisy quantum memories is done in parallel. In a one-hop link, distillation can be performed either as soon as two entangled states have arrived (\textsf{Distill-ASAP}), or it can be delayed until the end (\textsf{Distill-ALAP}). A simpler alternative is to discard the older pair (\textsf{Discard-Oldest}). We analytically derive the expected output fidelity, the success probability, and the weighted coherent information for these three strategies, and we find that in the low memory coherence time regime, \textsf{Discard-Oldest} achieves higher expected output fidelity, while in the high coherence time regime \textsf{Distill-ALAP} achieves the highest expected fidelity and, over most of the deadline range, the highest weighted coherent information, at the expense of a lower success probability. In a two-hop chain, entanglement swapping is also necessary, and we can give priority to either distillation or swapping. This yields seven total strategies: (1) \textsf{Distill-ASAP-then-Swap-ASAP}, (2) \textsf{Distill-ASAP-then-Swap-ALAP}, (3) \textsf{Distill-ALAP-then-Swap-ALAP}, (4) \textsf{Swap-ASAP-then-Distill-ASAP}, (5) \textsf{Swap-ASAP-then-Distill-ALAP}, (6) \textsf{Swap-ALAP-then-Distill-ALAP}, as well as (7) \textsf{Discard-Oldest-then-Swap}. For two-hop chains, a purely theoretical analysis will not be sufficiently helpful to elucidate the behavior of the complicated multi-parameter system, and we use Monte Carlo simulation. In this setting, our results show that deferring distillation to the end of the time window gives the highest weighted coherent information, with the leading strategy depending on the deadline: \textsf{Distill-ALAP-then-Swap-ALAP} leads at short deadlines and \textsf{Swap-ASAP-then-Distill-ALAP} leads as the deadline grows. These results clarify how memory decoherence affects the timing of operations and provide insights into the link-level guiding principles that may be instructive for larger-scale repeater architectures.

The remainder of this paper is organized as follows. In Sec.~\ref{sec:preliminaries}, we introduce the network model, noise model, distillation and swapping operations, detailed descriptions of strategies, and figures of merit. In Sec.~\ref{sec:one-hop}, we present the analytical treatment for one-hop links, including closed-form expressions for fidelity, success probability, and coherent information, as well as Monte Carlo simulation of fidelity and coherent information distributions. In Sec.~\ref{sec:two-hop}, we present the Monte Carlo analysis for two-hop chains and compare all seven strategies across a range of generation rates and coherence times. We conclude in Sec.~\ref{sec:conclusions} with a summary of findings and directions for future work.

\section{Preliminaries}\label{sec:preliminaries}
\subsection{Network Model}

\begin{figure}
    \centering
    \resizebox{\linewidth}{!}{%
    \begin{tikzpicture}[>=Latex, font=\small,
      node/.style={circle, fill=blue!70!black, minimum size=6pt, inner sep=0pt},
      src/.style={draw=black, fill=green!25!white, minimum width=1.15cm, minimum height=0.55cm, inner sep=1pt, font=\footnotesize},
      ens/.style={draw=blue!45!black, fill=blue!12!white, rounded corners=10pt}
    ]

    % Panel (a): multiplexed one-hop link
    \begin{scope}[shift={(3.25,0)}]
      \node[node] (aL1) at (0,0.55) {};
      \node[node] (aL2) at (0,-0.55) {};
      \node[node] (aR1) at (3.2,0.55) {};
      \node[node] (aR2) at (3.2,-0.55) {};
      \node[src] (aS1) at (1.6,0.55) {source};
      \node[src] (aS2) at (1.6,-0.55) {source};
      \draw (aL1)--(aS1); \draw (aS1)--(aR1);
      \draw (aL2)--(aS2); \draw (aS2)--(aR2);
      \begin{scope}[on background layer]
        \node[ens, fit=(aL1)(aL2), inner sep=6pt] (Aell) {};
        \node[ens, fit=(aR1)(aR2), inner sep=6pt] (Bell) {};
      \end{scope}
      \node[above=1pt of aL1, xshift=-10pt] {\textbf{A}};
      \node[above=1pt of aR1, xshift=10pt] {\textbf{B}};
      \node[above=2pt of aS1] {$t_1$};
      \node[below=2pt of aS2] {$t_2$};
      \node[below=1.2cm] at (1.6,-0.55) {(a) Multiplexed one-hop link.};
    \end{scope}

    % Panel (b): two-hop repeater chain
    \begin{scope}[shift={(1.50,-4.0)}]
      \node[node] (bA1) at (0,0.55) {};
      \node[node] (bA2) at (0,-0.55) {};
      \node[node] (bC1) at (3.0,0.55) {};
      \node[node] (bC2) at (3.7,0.55) {};
      \node[node] (bC3) at (3.0,-0.55) {};
      \node[node] (bC4) at (3.7,-0.55) {};
      \node[node] (bB1) at (6.7,0.55) {};
      \node[node] (bB2) at (6.7,-0.55) {};
      \node[src] (bSL1) at (1.5,0.55) {source};
      \node[src] (bSL2) at (1.5,-0.55) {source};
      \node[src] (bSR1) at (5.2,0.55) {source};
      \node[src] (bSR2) at (5.2,-0.55) {source};
      \draw (bA1)--(bSL1); \draw (bSL1)--(bC1);
      \draw (bA2)--(bSL2); \draw (bSL2)--(bC3);
      \draw (bC2)--(bSR1); \draw (bSR1)--(bB1);
      \draw (bC4)--(bSR2); \draw (bSR2)--(bB2);
      \draw[densely dotted] (bC1)--(bC2);
      \draw[densely dotted] (bC3)--(bC4);
      \draw[densely dotted] (bC1)--(bC4);
      \draw[densely dotted] (bC3)--(bC2);
      \begin{scope}[on background layer]
        \node[ens, fit=(bA1)(bA2), inner sep=6pt] {};
        \node[ens, fit=(bC1)(bC2)(bC3)(bC4), inner sep=6pt] (Cell) {};
        \node[ens, fit=(bB1)(bB2), inner sep=6pt] {};
      \end{scope}
      \node[above=1pt of bA1, xshift=-10pt] {\textbf{A}};
      \node[above=1pt of bB1, xshift=10pt] {\textbf{B}};
      \node[above=8pt] at (Cell.north) {\textbf{C}};
      \node[above=2pt of bSL1] {$t_1$};
      \node[below=2pt of bSL2] {$t_2$};
      \node[above=2pt of bSR1] {$t_3$};
      \node[below=2pt of bSR2] {$t_4$};
      \node[below=1.2cm, align=center] at (3.35,-0.55)
        {(b) Two-hop repeater chain. Cross-channel entanglement\\ swapping is assumed possible as shown by the dashed lines.};
    \end{scope}

    % Panel (c): event dynamics
    \begin{scope}[shift={(0,-11.0)}]
      \def\xend{10.2}
      \draw[->,thick] (-0.5,0) -- (\xend,0) node[right] {$t$};

      \coordinate (t0)      at (-0.1,0);
      \coordinate (a1)      at (1.1,0);
      \coordinate (a2)      at (2.5,0);
      \coordinate (b1tswap) at (4.0,0);
      \coordinate (b2)      at (5.6,0);
      \coordinate (TT)      at (9.0,0);

      \foreach \p in {t0,a1,a2,b1tswap,b2,TT}{\draw[thick] ($(\p)+(0,0.1)$) -- ($(\p)+(0,-0.1)$);}
      \node[below=3pt] at (t0)      {$t=0$};
      \node[below=3pt] at (a1)      {$a_1$};
      \node[below=3pt] at (a2)      {$a_2$};
      \node[below=3pt, align=center] at (b1tswap) {$b_1$\\[-1pt]$=t_{\text{swap}}$};
      \node[below=3pt] at (b2)      {$b_2$};
      \node[below=3pt] at (TT)      {$T$};

      \draw[densely dashed] ($(b1tswap)+(0,0.35)$) -- ($(TT)+(0,0.35)$);
      \node[font=\footnotesize] at ($(b1tswap)!0.5!(TT)+(0,0.62)$) {swapped pair idles};

      \draw[->,thick] ($(a2)+(0,1.0)$) -- ($(a2)+(0,0.14)$);
      \node[font=\footnotesize] at ($(a2)+(0,1.28)$) {\emph{distill A}};

      \draw[->,thick] ($(b1tswap)+(0,2.6)$) -- ($(b1tswap)+(0,0.14)$);
      \node at ($(b1tswap)+(0,2.88)$) {\emph{swap}};

      \draw[->,thick] ($(TT)+(0,2.6)$) -- ($(TT)+(0,0.14)$);
      \node at ($(TT)+(0,2.88)$) {\emph{consume}};
      \node[below=1.25cm, align=center, text width=9.5cm] at (4.85,0) {(c) Temporal dynamics of the \textsf{Distill-ASAP-then-Swap-ASAP} strategy.};
    \end{scope}

    \end{tikzpicture}%
    }
    \caption{Illustrations of (a) a one-hop network and (b) a two-hop network. The entanglement generation process is attempted continuously, and the labels $t_i$ on the links denote the times at which the initial entangled states are stored into the memory. (c) Illustration of the temporal dynamics of the \textsf{Distill-ASAP-then-Swap-ASAP} (D-ASAP-S-ASAP) strategy: the swap occurs at $t_{\text{swap}}=\max(a_1,b_1)$; a segment distills before the swap only if its second copy has already arrived by $t_{\text{swap}}$, as shown here for link A ($a_2 \le t_{\text{swap}}$), while link B's second copy arrives too late ($b_2 > t_{\text{swap}}$) to be distilled. The swapped pair then idles until it is consumed at the final time $T$.}
    \label{fig:diagrams}
\end{figure}

Near-term networks typically suffer from probabilistic, low-rate entanglement generation. We assume that the source generates raw isotropic states $\rho_0$, whose fidelity with respect to the Bell state $\phi_+=\ket{\phi_+}\bra{\phi_+}$, where $\ket{\phi_+}=(\ket{00}+\ket{11})/\sqrt{2}$, is $F_0$. That is,
\begin{align}
    \rho_0 = F_0\phi_++(1-F_0)\frac{\mathbb{I}-\phi_+}{3}.
\end{align}
We will henceforth assume $F_0 = 0.9$ unless otherwise specified, representing imperfect entanglement generation typical of near-term quantum sources~\cite{krutyanskiy2023entanglement}. 

The entanglement generation is continuously attempted with heralded success at a constant rate $\lambda$. The probability density for the arrival time of the first generated entangled state for each individual channel is thus given by the exponential distribution $p(t)=\lambda e^{-\lambda t}$. Upon a heralded success, it is stored into a pair of quantum memories. Once stored, an entangled state is not overwritten unless otherwise specified. We assume that quantum memories are affected by depolarizing noise, modeled by the quantum channel $\mc{N}(\Delta t)=e^{-\Delta t/\tau}\mathrm{id} + (1-e^{-\Delta t/\tau})\mc{D}$, where $\tau$ is the coherence time of the memory, $\mathrm{id}$ is the identity channel, and $\mc{D}$ is the completely depolarizing channel $\mc{D}(X)=\Tr(X) \mathbb{I}/2$. Therefore, after being stored in the quantum memories for time $\Delta t$, the entangled state becomes (note that the memory noise will affect both subsystems) $\rho(\Delta t) = [\mc{N}(\Delta t)\otimes\mc{N}(\Delta t)](\rho_0) = e^{-2\Delta t/\tau}\left(F_0\phi_+ + (1-F_0)\frac{\mathbb{I}-\phi_+}{3}\right) + (1-e^{-2\Delta t/\tau})\frac{\mathbb{I}}{4}$. The fidelity of this state is
\begin{align}
    F(\Delta t) = \frac{1}{4} + \left(F_0-\frac{1}{4}\right)e^{-2\Delta t/\tau}.
\end{align}

Once two entangled states are generated on the same segment and stored into quantum memories, we can opt to perform entanglement distillation to extract one high-fidelity entangled state from the two stored states. In the present paper, we will consider the conventional CNOT-based entanglement distillation protocol~\cite{bennett1996purification,deutsch1996quantum}. It is the optimal 2-to-1 bilocal Clifford protocol for identical Werner states~\cite{jansen2022enumerating}. While the CNOT protocol has not yet been proven to be optimal in the most general sense, there is already strong analytical and numerical evidence of its optimality~\cite{rozpkedek2018optimizing,preti2022optimal}. From a practical perspective, it is also arguably the most experiment-friendly entanglement distillation protocol given its realization in various physical systems~\cite{pan2001entanglement,pan2003experimental,hu2021long,ecker2021experimental,reichle2006experimental,kalb2017entanglement,yan2022entanglement}. Given two isotropic states with fidelities $F_1$ and $F_2$, the protocol succeeds with probability 
\begin{align}
    P_\text{dist}(F_1,F_2) =& F_1 F_2 + \frac{1}{3} \Big[F_1 (1-F_2) + F_2 (1-F_1)\Big]\notag\\
    &+ \frac{5}{9} (1-F_1)(1-F_2),
\end{align}
and the output state conditioned on success has fidelity
\begin{align}
    F_\text{dist}(F_1,F_2) = \frac{F_1 F_2 + \frac{1}{9}(1-F_1)(1-F_2)}{P_\text{dist}(F_1,F_2)}.
\end{align}
We also assume that the state after distillation is twirled into an isotropic state.

In two-hop repeater chains, we need to perform entanglement swapping to connect entangled states into a long-range entangled state across two links. We reiterate that we allow entanglement swapping to be performed cross-channel (see Fig.~\ref{fig:diagrams}). Given two isotropic states with fidelities $F_1$ and $F_2$, the state after swapping is an isotropic state with fidelity
\begin{align}
    F_\text{swap}(F_1,F_2) = F_1F_2+\frac{1}{3}(1-F_1)(1-F_2).
\end{align}
In matter-qubit systems, Bell state measurements can be implemented by rotating the measurement basis into the computational basis, and therefore we assume that swapping succeeds with unit probability. In linear-optical systems, on the other hand, Bell state measurements succeed with 1/2 probability~\cite{calsamiglia2001maximum}, which can be enhanced to $1-1/2^n$ by using $2^n-2$ ancillary photons~\cite{grice2011arbitrarily}. In this work, we assume for simplicity that the swaps succeed with unit probability. Similar analyses can also be carried out for probabilistic entanglement swapping. We will evaluate the quality of the final end-to-end entangled state at a predetermined end time $T$, which may be regarded as the time when the entangled state is consumed. An illustration of the operation times is provided in Fig.~\ref{fig:diagrams}.

Here, a few very natural questions arise. Should we perform entanglement distillation on individual segments before performing swapping to connect the segments, or should we perform swapping before performing distillation on two long-range entangled states? Should we perform these operations as soon as we have obtained a sufficient number of states and let the final state idle in the memories, or should we perform the operations at the end to minimize idle time? It is precisely our goal to address these questions. In the next subsection, we will provide a more detailed description of each strategy.

\subsection{Strategy description}\label{sec:protocol-description}

A key contribution of our work is the explicit modeling of timing policies that determine when and how to perform distillation and swapping operations in the presence of probabilistic link generation and memory decoherence. The choice of operational policy represents a fundamental design decision in quantum repeater networks, balancing the benefits of immediate operation against the risks of accumulated decoherence.

All strategies below are \emph{causal}: every operation time is chosen using only the arrivals that have already occurred at that moment. In particular, an ASAP operation is never deferred in order to wait for a pair that has not yet been generated.

\subsubsection{One-Hop Networks}

For one-hop scenarios where only distillation is performed (no swapping), we investigate three distinct operational strategies:

\begin{itemize}
    \item \textsf{1. Distill-ASAP.} Distill as soon as two states are stored in memory. If only one state arrives, the state is returned as the output state. If two states are stored and distillation succeeds, the distilled state remains idle in memory and is returned as output at the end. If distillation fails or if no state arrives, the strategy fails.

    \item \textsf{2. Distill-ALAP.} Distill only at the end. If only one state arrives, the state is returned as the output state. If at the end two states are stored and distillation succeeds, the distilled state is returned as output. If distillation fails or if no state arrives, the strategy fails.

    \item \textsf{3. Discard-Oldest.} The strategy succeeds if at least one state is stored at the end. If two states are stored, the older state is discarded. 
\end{itemize}

\subsubsection{Two-Hop Networks}
For two-hop scenarios, a crucial problem is the timing of distillation and swapping. If we opt to perform distillation first, we still have the option to choose when to perform distillation and swapping. This yields three possible strategies:

\begin{itemize}
    \item 1. \textsf{Distill-ASAP-then-Swap-ASAP (D-ASAP-S-ASAP)}. Distillation is prioritized over swapping. The swap is performed at the first instant $t_s$ at which both segments hold at least one pair, that is, $t_s=\max\{a_1,b_1\}$ where $a_1$ and $b_1$ are the first arrivals on the two segments; a segment whose second pair arrives later contributes its single stored pair, and the later arrival is not waited for. The swapped state then idles until $T$.

    \item 2. \textsf{Distill-ASAP-then-Swap-ALAP (D-ASAP-S-ALAP).} Each segment distills as soon as its second pair arrives, the distilled state is stored, and the swap is performed at the end of the time window~$T$.

    \item 3. \textsf{Distill-ALAP-then-Swap-ALAP (D-ALAP-S-ALAP).} Both the per-segment distillation and the swap are performed at the end of the time window~$T$.
\end{itemize}

We have three similar strategies when we opt to perform swapping before distillation:
\begin{itemize}
    \item 4. \textsf{Swap-ASAP-then-Distill-ASAP (S-ASAP-D-ASAP)};
    \item 5. \textsf{Swap-ASAP-then-Distill-ALAP (S-ASAP-D-ALAP)};
    \item 6. \textsf{Swap-ALAP-then-Distill-ALAP (S-ALAP-D-ALAP)}.
\end{itemize}
Since entanglement distillation requires two copies of an end-to-end entangled state held jointly by the two end parties, a second copy to distill against can only be produced by a second full swap, which in turn requires a second elementary pair to have arrived on both segments before $T$. A single leftover elementary pair on only one segment is not, by itself, a valid input to the distillation step. For the swap-ASAP variants both swaps are performed as soon as their own inputs exist, at $t_{s,1}=\max\{a_1,b_1\}$ and $t_{s,2}=\max\{a_2,b_2\}$, respectively; the distillation of the two end-to-end copies is then either performed as soon as the second copy exists (D-ASAP) or deferred to $T$ (D-ALAP). Again, it is possible to just discard the older state instead of performing distillation.
\begin{itemize}
    \item 7. \textsf{Discard-Oldest-then-Swap (Discard-Swap)}. The strategy succeeds if at least one state is stored on each segment. At time $T$, the one-hop \textsf{Discard-Oldest} is performed on each link. Then entanglement swapping is performed.
\end{itemize}

\subsection{Strategy evaluation}\label{sect:protocol-evaluation}
To evaluate the performance of a strategy $\mc{S}$, we will use three figures of merit. A natural one is the expected fidelity conditioned on success
\begin{align}
    \bar{F}_{\mc{S}}=\mathbb{E}[F|\text{success}].
\end{align}
This quantity aptly captures the quality of the output entangled state, but it fails to take into account the strategy's success probability. A complementary figure of merit is therefore the overall success probability at the end of the time window $T$, which accounts for all intermediate probabilistic steps, including any distillation attempts that may occur
\begin{align}
    P_{\mc{S}} = \mathrm{Pr}(\text{success}).
\end{align}
In the Monte Carlo estimates below, both quantities are computed from the same per-realization success weights: writing $w_j$ for the product of the success probabilities of the distillations actually attempted in realization $j$ and $F_j$ for the corresponding conditional output fidelity, we use $P_\mc{S}=\frac{1}{N}\sum_j w_j$ and $\bar F_\mc{S}=\sum_j w_j F_j/\sum_j w_j$. Averaging candidate fidelities with unit weight would describe a different, hypothetical process.

Furthermore, we can attempt to combine these two quantities into a single measure that quantifies the performance. To this end, we will consider the coherent information of the expected output state
\begin{align}
    R_{\mc{S}} = P_{\mc{S}}\cdot \max\left\{I_c(\mathbb{E}[\rho_\text{out}|\text{success}]), 0\right\},
\end{align}
where $I_c(\rho^{AB})=S(B)-S(AB)$ is the coherent information of state $\rho$. We adopt the coherent information as a performance measure because of its operational significance. Coherent information $I_c(\rho)$ represents the rate at which maximally entangled states can be extracted from the state $\rho$ in the asymptotic regime using the hashing protocol~\cite{devetak2005distillation}. Note that coherent information may be negative, in which case the entanglement distillation rate is effectively zero. Hence, we have taken the maximum of $I_{c}$ and 0 in our definition of $R_\mc{S}$. Since the strategies succeed with probability $P_\mc{S}$, the coherent information weighted by $P_\mc{S}$ is a meaningful measure that quantifies the amount of entanglement that can be extracted from the output, taking into account both the quality of the output and the success probability of the strategy. For an isotropic state with fidelity $F$, its coherent information is $I_c(\rho)=1-h_2(F)-(1-F)\log_2 3$, where $h_2(x)=-x\log_2 x-(1-x)\log_2(1-x)$ is the binary entropy function. Therefore,
\begin{align}
    R_\mc{S} = P_\mc{S} \cdot \max\left\{[1-h_2(\bar{F}_\mc{S})-(1-\bar{F}_\mc{S})\log_2 3], 0\right\}. \label{eq:coh-info-expression}
\end{align}

\section{Analysis for One-Hop Strategies}\label{sec:one-hop}

\begin{figure*}
    \centering
    \includegraphics[width=0.7\linewidth]{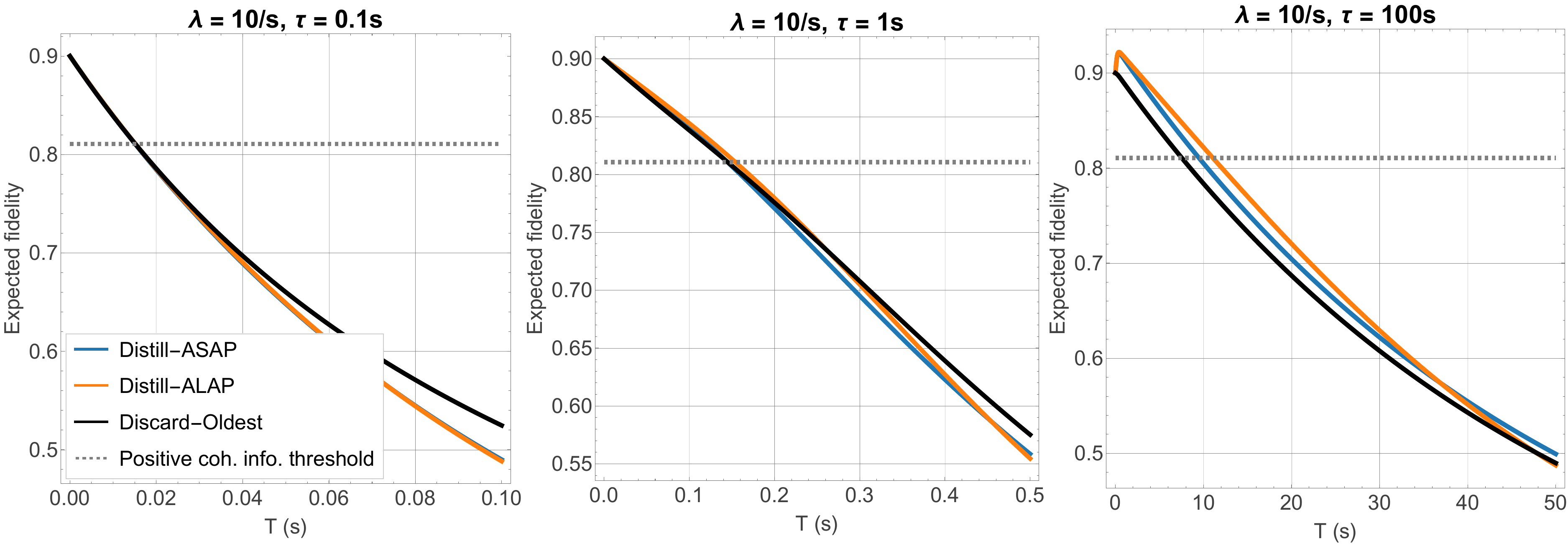}
    \caption{The expected fidelity $\bar{F}$ for one-hop strategies with $\lambda=10/\mathrm{s}$ and $\tau=0.1, 1, 100\,\mathrm{s}$.}
    \label{fig:one-link-fidelity}
\end{figure*}

\begin{figure*}
    \centering
    \includegraphics[width=0.7\linewidth]{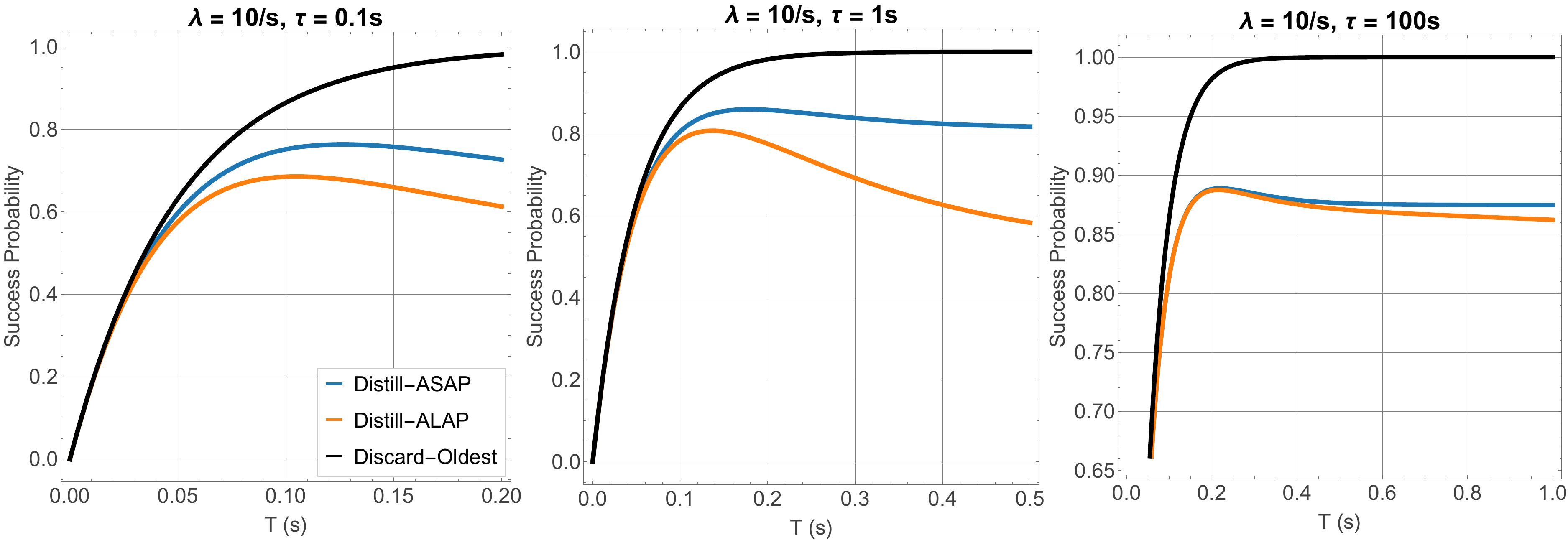}
    \caption{The success probability $\mathrm{Pr}(\text{success})$ for one-hop strategies with $\lambda=10/\mathrm{s}$ and $\tau=0.1, 1, 100\,\mathrm{s}$.}
    \label{fig:one-link-psucc}
\end{figure*}

\begin{figure*}
    \centering
    \includegraphics[width=0.7\linewidth]{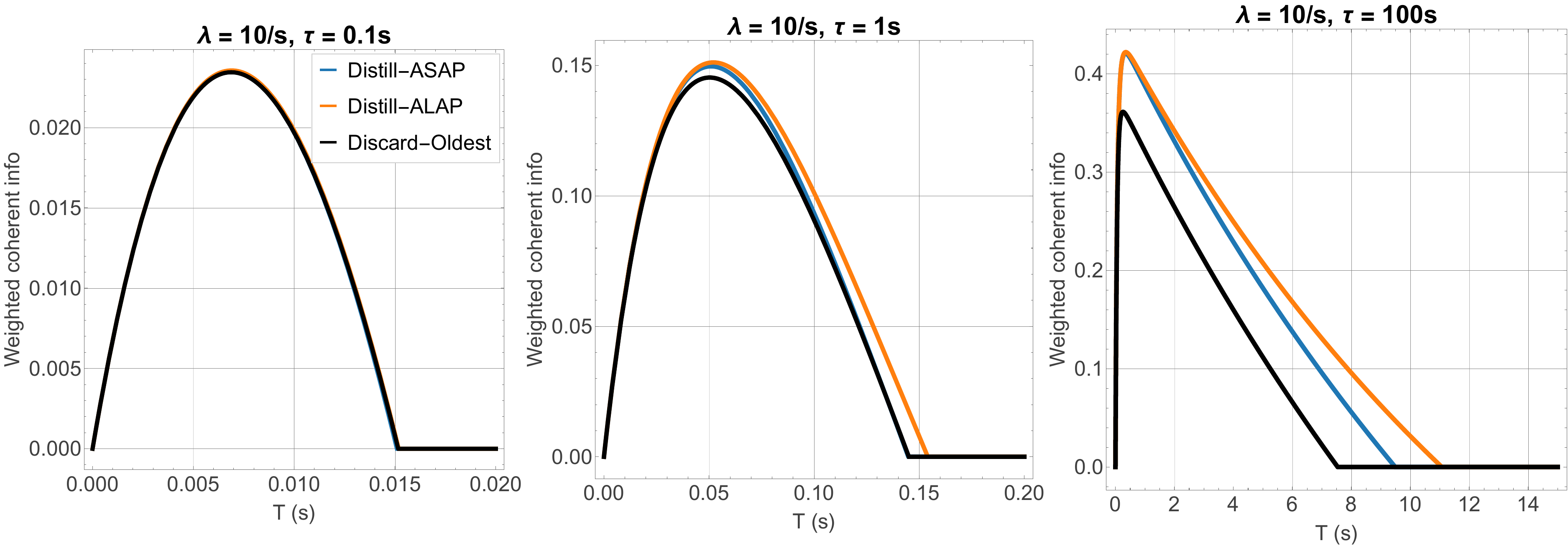}
    \caption{The weighted coherent information $R$ for one-hop strategies with $\lambda=10/\mathrm{s}$ and $\tau=0.1, 1, 100\,\mathrm{s}$.}
    \label{fig:one-link-Ic}
\end{figure*}

We begin with our analysis of entanglement processing on a one-hop chain. The performance of the one-hop strategies is sufficiently simple to be treated analytically.

\subsection{Discard-Oldest}

\textsf{Discard-Oldest} succeeds if at least one entangled state has been stored in either of the two memory pairs available at time $T$. This happens with probability
\begin{align}
    P_\textsf{Disc} = 1 - \left(\int_T^\infty dt\ p(t)\right)^2 = 1-e^{-2\lambda T}.
\end{align}

If the first source generates entanglement at time $t_1\in[0,T]$ while the second source does not (in other words, the second source generates entanglement at $t_2\in[T,\infty)$), then the fidelity of that state at time $T$ is given by $F(T-t_1)$. If both $t_1,t_2\in[0,T]$, the fidelity of the newer pair at $T$ is $F(T-\max\{t_1,t_2\})$. The expected output fidelity is therefore
\begin{align}
    \bar{F}_\textsf{Disc} = \frac{I_1+I_\textsf{Disc}}{P_\textsf{Disc}},
\end{align}
where 
\begin{align}
    I_1 = 2\int_0^T dt_1\int_T^\infty dt_2\ p(t_1)p(t_2)F(T-t_1), \label{eq:I1}
\end{align}
and
\begin{IEEEeqnarray}{rCl}
    I_\textsf{Disc} = \int_0^T dt_1 \int_0^T dt_2\ p(t_1)p(t_2)F(T-\max\{t_1,t_2\}). \IEEEeqnarraynumspace
\end{IEEEeqnarray}
The weighted coherent information of the expected output state $R_\textsf{Disc}$ can then be readily calculated with Eq.~\ref{eq:coh-info-expression}.

\subsection{Distill-ASAP}
The probability that \textsf{Distill-ASAP} succeeds is
\begin{align}
    P_\textsf{D-ASAP} 
    &= \mathrm{Pr}(n=1)+\mathrm{Pr}(n=2,\text{distillation success}) \notag\\
    &= 2\int_0^T dt_1\int_T^\infty dt_2 p(t_1)p(t_2) \notag\\
    &\quad+ \int_0^T dt_1\int_0^T dt_2 p(t_1)p(t_2)P_\text{dist}(t_1,t_2),
\end{align}
where $n$ is the number of successfully generated entangled states, and $P_\text{dist}$ is given by
\begin{align}
    P_\text{dist}(t_1,t_2) = P_\text{dist}\big(F(0),F(|t_2-t_1|)\big).
\end{align}

Again, if the first source generates entanglement at $t_1\in[0,T]$ and the second source generates entanglement at $t_2\in[T,\infty)$, the fidelity of the output state is given by $F(T-t_1)$. If $t_1,t_2\in[0,T]$, distillation is performed ASAP at time $\max\{t_1,t_2\}$. If distillation is successful, the fidelity of the distillation output immediately after distillation is
\begin{align}
    F_\text{dist}(t_1,t_2) = F_\text{dist}\big(F(0),F(|t_2-t_1|)\big).
\end{align}
The distilled state idles in the memories for time $T-\max\{t_1,t_2\}$. Then the fidelity of the state at time $T$ is $\frac{1}{4}+\left(F_\text{dist}(t_1,t_2)-\frac{1}{4}\right)e^{-2(T-\max\{t_1,t_2\})/\tau}$. Altogether, the expected output fidelity is
\begin{align}
    \bar{F}_\textsf{D-ASAP} = \frac{I_1+I_\textsf{D-ASAP}}{P_\textsf{D-ASAP}},
\end{align}
where $I_1$ has been given in Eq.~\ref{eq:I1}, and
\begin{IEEEeqnarray}{rCl}
    I_\textsf{D-ASAP} &=& \int_0^T dt_1 \int_0^T dt_2\ p(t_1)p(t_2)P_\text{dist}(t_1,t_2) \notag\\
    &&\times \left[\frac{1}{4}+ \left(F_\text{dist}(t_1,t_2)-\frac{1}{4}\right)e^{-2(T-\max\{t_1,t_2\})/\tau}\right]. \IEEEeqnarraynumspace
\end{IEEEeqnarray}
We can similarly calculate the weighted coherent information of the expected output with Eq.~\ref{eq:coh-info-expression}.

\subsection{Distill-ALAP}
The success probability of \textsf{Distill-ALAP} is
\begin{align}
    &P_\textsf{D-ALAP} = \mathrm{Pr}(n=1)+\mathrm{Pr}(n=2,\text{distillation success}) \notag\\
    &~~~~~~~~~~ = 2\int_0^T dt_1\int_T^\infty dt_2 p(t_1)p(t_2) \notag\\
    &~~~~~~~~~~\quad + \int_0^T dt_1\int_0^T dt_2 p(t_1)p(t_2)P_\text{dist}(t_1,t_2), \label{eq:Distill-ALAP-prob-succ}
\end{align}
where $n$ is again the number of successfully generated entangled states. Here, we have used $P_\text{dist}(t_1,t_2)$ to denote the probability that entanglement distillation succeeds if the two entangled states are generated at time $t_1$ and $t_2$. For \textsf{Distill-ALAP} specifically, it is given by $P_\text{dist}(t_1,t_2)=P_\text{dist}\big(F(T-t_1),F(T-t_2)\big)$. 

As before, if the first source generates entanglement at time $t_1\in[0,T]$ while the second entangled state's arrival time is at $t_2\in[T,\infty)$, then the fidelity of that state at time $T$ is given by $F(T-t_1)$. If both $t_1,t_2\in[0,T]$ and the distillation performed at time $T$ succeeds, the fidelity of the distillation output is 
\begin{align}
    F_\text{dist}(t_1,t_2) = F_\text{dist}\big(F(T-t_1),F(T-t_2)\big).
\end{align}
So,
\begin{align}
    \bar{F}_\textsf{D-ALAP} = \frac{I_1+I_\textsf{D-ALAP}}{P_\textsf{D-ALAP}}, \label{eq:D-ALAP-fidelity}
\end{align}
where $I_1$ was given in Eq.~\ref{eq:I1} and
\begin{IEEEeqnarray}{rCl}
    I_\textsf{D-ALAP} &=& \int_0^T dt_1 \int_0^T dt_2 p(t_1)p(t_2) P_\text{dist}(t_1,t_2)F_\text{dist}(t_1,t_2). \IEEEeqnarraynumspace
\end{IEEEeqnarray}
Again, the weighted coherent information of the expected output is given according to Eq.~\ref{eq:coh-info-expression}.

\subsection{Performance Comparison}\label{sect:one-link-comparison}

The expressions set up above can be analytically evaluated using Mathematica. The Mathematica notebooks used for this evaluation, together with the Python scripts used for the Monte Carlo simulations in the following sections, are available at~\cite{github}, and the closed-form results are given in Appendix~\ref{sect:analytical-expressions}.

Fig.~\ref{fig:one-link-fidelity} plots the analytical expression for the expected output fidelity for varying memory coherence times. When the coherence time is small ($\tau=0.1\ \mathrm{s}$ or $\tau=1\ \mathrm{s}$), \textsf{Discard-Oldest} achieves higher expected output fidelity than \textsf{Distill-ASAP} and \textsf{Distill-ALAP} once the deadline is long enough for the first pair to have decohered appreciably. This is because in this regime, the first entangled pair has decohered enough that the fidelity of the output pair is in fact lower than that of the second pair alone. Therefore, discarding the older pair is more advantageous. At the shortest deadlines the three curves remain close, since little decoherence has accumulated either way. As the coherence time increases, distillation-based strategies begin to exhibit an advantage over \textsf{Discard-Oldest}. For $\tau=100\ \mathrm{s}$, \textsf{Distill-ALAP} achieves the highest expected output fidelity, followed by \textsf{Distill-ASAP}, with \textsf{Discard-Oldest} lowest, until the longest deadlines. This shows that in the high-coherence regime the advantage gained by distillation more than compensates for the memory decoherence.

The success probabilities are plotted in Fig.~\ref{fig:one-link-psucc}. The success probability of \textsf{Discard-Oldest} approaches 1 when $T$ is large. This is because the probability that at least one entangled state is generated approaches unity. The success probability of \textsf{Distill-ASAP} approaches a constant that is fixed by $F_0$, $\lambda$, and $\tau$. Specifically, when $T$ is sufficiently large, two states will be generated in time $T$ with high probability. The probability that distillation succeeds is approximately
\begin{align}
    P &\approx P_\text{dist}(F_0,F_0) \qquad\text{if}\ \lambda\tau\gg1 \notag\\
    &= F_0^2 + \frac{2}{3}F_0(1-F_0) + \frac{5}{9}(1-F_0)^2.
\end{align}
For $F_0=0.9$, this evaluates to approximately 0.876, which matches the behavior we observe for $\lambda=10/\mathrm{s}$ and $\tau=100\ \mathrm{s}$. On the other hand, the success probability of \textsf{Distill-ALAP} rises initially as more time allows both pairs to arrive, but eventually decreases as the original states are stored in the memories for a longer time, resulting in lower fidelities and therefore lower distillation success probability. 

The weighted coherent information (Fig.~\ref{fig:one-link-Ic}) exhibits a peak due to the increasing success probability and decreasing expected output fidelity. In the low coherence time regime the three strategies achieve essentially the same weighted coherent information, the curves being indistinguishable at $\tau=0.1\ \mathrm{s}$. At $\tau=1\ \mathrm{s}$ the strategies separate only slightly, with \textsf{Distill-ALAP} marginally ahead of \textsf{Distill-ASAP} near the peak and on the falling edge, sustaining positive $R$ to a longer deadline. At $\tau=100\ \mathrm{s}$ the ordering is unambiguous, with \textsf{Distill-ALAP} highest, \textsf{Distill-ASAP} intermediate and \textsf{Discard-Oldest} lowest, and \textsf{Distill-ALAP} retaining positive $R$ well past the deadline at which \textsf{Discard-Oldest} has fallen to zero. The conclusions we draw in this section represent a general trend and apply to different parameter choices.

\subsection{Fidelity Distributions}
\begin{figure}
    \centering
    \subfloat[]{\includegraphics[width=0.5\linewidth]{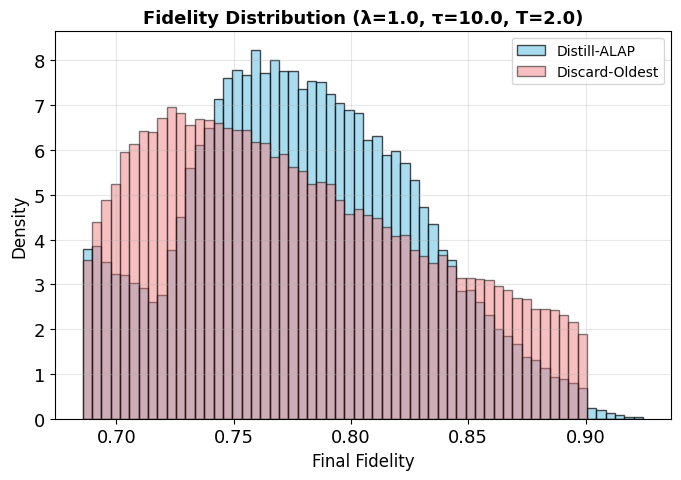}
    \label{fig:one-link-fidelity-distribution}}
    \subfloat[]{
    \includegraphics[width=0.5\linewidth]{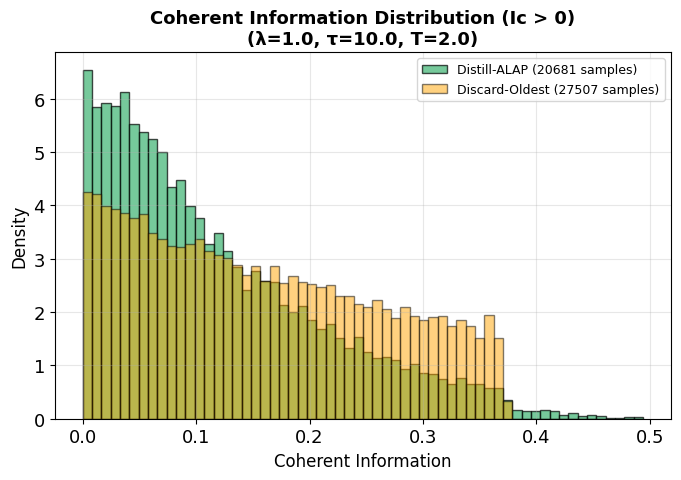} \label{fig:one-link-Ic-distribution}
    }
    \caption{(a) Distribution of final fidelities for \textsf{Distill-ALAP} (mean $\bar{F}\approx0.780$, standard deviation $\approx0.048$) and \textsf{Discard-Oldest} (mean $\bar{F}=0.776$, standard deviation $0.056$) at fixed parameters ($\lambda=1.0\ \mathrm{s}^{-1}$, $\tau=10.0\ \mathrm{s}$, $T=2.0\ \mathrm{s}$, $10^5$ attempted samples). (b) Distribution of coherent information conditioned on $I_c > 0$ at the same parameters. Of $10^5$ attempted samples, $20{,}681$ (\textsf{Distill-ALAP}) and $27{,}507$ (\textsf{Discard-Oldest}) successful outputs have positive $I_c$.}
\end{figure}

In addition to analytical derivations of the expected fidelity, success probability, and coherent information, we also obtain the full statistical distribution of fidelity and coherent information using Monte Carlo simulation at fixed parameters ($\lambda=1\ \text{s}^{-1}$, $T=2\ \text{s}$, $\tau=10\ \text{s}$, $10^5$ samples). Failed distillation attempts produce no output state and are excluded from these distributions. The results are plotted in Figs.~\ref{fig:one-link-fidelity-distribution} and~\ref{fig:one-link-Ic-distribution}.

Figure~\ref{fig:one-link-fidelity-distribution} shows the output fidelity distributions for \textsf{Distill-ALAP} and \textsf{Discard-Oldest}. Neither distribution is a simple unimodal shape: the \textsf{Distill-ALAP} distribution is a mixture, whose lower component comes from realizations in which only one pair arrived and is returned undistilled, and whose upper component comes from realizations in which two pairs arrived and distillation succeeded. For \textsf{Discard-Oldest}, the distribution rises from $F(T) \approx 0.686$ (the fidelity of a pair generated at $t=0$ and stored for the full duration $T$) and is truncated sharply at the initial source fidelity $F_0 = 0.9$, since the newest available pair can at best arrive just before $T$ with negligible storage time. 

This hard upper bound is a structural feature of the \textsf{Discard-Oldest} strategy. The \textsf{Distill-ALAP} distribution, by contrast, extends beyond $F_0 = 0.9$, reaching up to $F \approx 0.926$. This tail arises because distillation can increase the fidelity above the source fidelity $F_0$ when both pairs arrive close to the end of the time window and suffer little decoherence: One can verify that $F_{\mathrm{dist}}(F_0, F_0) \approx 0.926 > F_0$. This is a direct signature of the fidelity-boosting capability of distillation and has no analogue in the \textsf{Discard-Oldest} strategy. The \textsf{Distill-ALAP} distribution is also narrower and shifted to higher fidelities on average (mean $\bar{F}\approx0.780$, standard deviation $\approx0.048$) versus \textsf{Discard-Oldest} (mean $\bar{F}=0.776$, standard deviation $0.056$), consistent with the analytical results in Fig.~\ref{fig:one-link-fidelity}.

The coherent information distributions in Fig.~\ref{fig:one-link-Ic-distribution} have a qualitatively different character from the fidelity distributions. For an isotropic state, coherent information is positive only when $F > F^{*}$, where $F^{*} \approx 0.811$ is the threshold at which $I_c(\rho) = 0$. Since the bulk of both fidelity distributions lies below this threshold, the majority of realizations contribute $I_c = 0$; Fig.~\ref{fig:one-link-Ic-distribution} therefore conditions on the minority of realizations with $I_c > 0$. The conditional distributions are both generally decreasing, and the values are just above zero, reflecting that even among the realizations with positive $I_c$, most fidelities lie only marginally above the threshold. 

Over the intermediate range of positive $I_c$, roughly $0.15$ to $0.37$, the \textsf{Discard-Oldest} distribution carries more probability density than \textsf{Distill-ALAP}, whose density falls off faster there. The maximum \emph{support}, however, runs the other way: the \textsf{Distill-ALAP} histogram extends to $I_c \approx 0.5$ while \textsf{Discard-Oldest} stops near $0.37$, since only distillation can push the output fidelity above $F_0$. Tail weight and maximum support therefore point in opposite directions here and should not be conflated. This arises because \textsf{Distill-ALAP} concentrates probability mass in a narrower fidelity range straddling $F^*$, whereas the broader \textsf{Discard-Oldest} distribution places more weight in the intermediate high-fidelity range above $F^*$. Nevertheless, the weighted coherent information $R_{\mathcal{S}} = P_{\mathcal{S}} \cdot \max\{I_c(\mathbb{E}[\rho_{\mathrm{out}}|\mathrm{success}]),0\}$, computed as $I_c$ of the mean output state weighted by the full strategy success probability (distinct from the per-sample distribution shown in Fig.~\ref{fig:one-link-Ic-distribution}), favors \textsf{Distill-ALAP} overall, as shown in Fig.~\ref{fig:one-link-Ic}.

\section{Analysis for Two-Hop Strategies}\label{sec:two-hop}
Building upon the one-hop foundation, we now analyze two-hop strategies that incorporate both distillation and swapping operations. The addition of swapping introduces new timing dependencies and operation ordering choices that significantly affect performance. Moreover, analytical methods for closed-form results become intractable, while Monte Carlo simulation is more convenient. On each segment we draw the arrival times of the two parallel sources from the exponential distribution $p(t)=\lambda e^{-\lambda t}$ and order them, so that the first and second arrivals on that segment are well defined, and we then track the fidelity dynamics under each strategy. The results are shown in Figs.~\ref{fig:double-link-fidelity-lam10},~\ref{fig:double-link-psucc-lam10}, and~\ref{fig:double-link-Ic-lam10}, where the expected output fidelity, success probability, and the weighted coherent information are plotted, respectively.

\subsection{Performance Comparison}

\begin{figure*}
    \centering
    \includegraphics[width=0.8\linewidth]{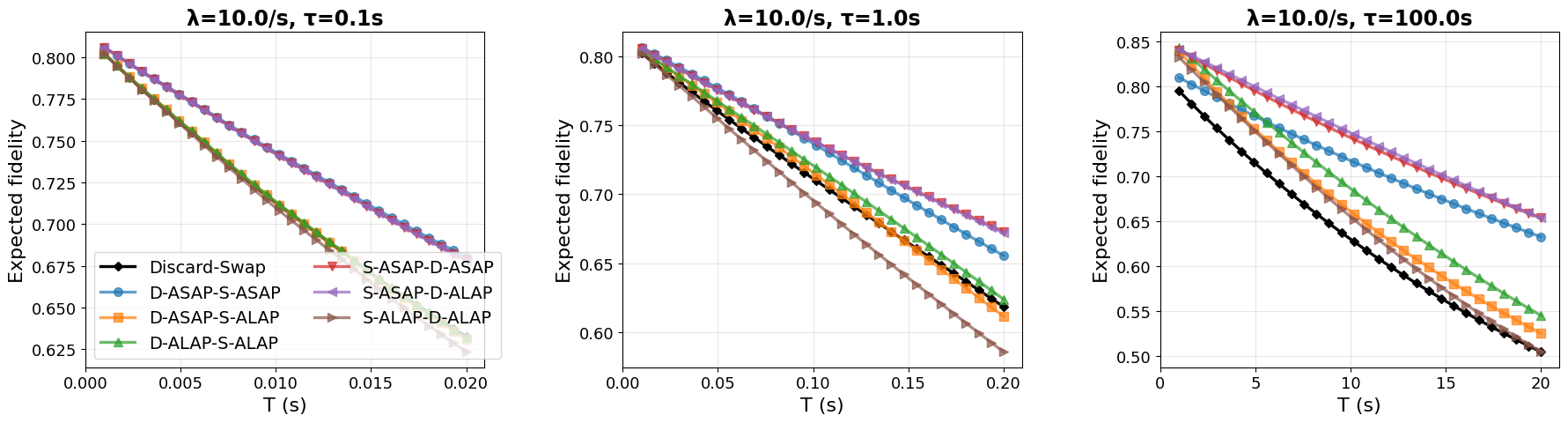}
    \caption{The expected fidelity $\bar{F}$ for two-hop strategies with $\lambda=10/\mathrm{s}$ and $\tau=0.1, 1, 100\,\mathrm{s}$.}
    \label{fig:double-link-fidelity-lam10}
\end{figure*}

\begin{figure*}
    \centering
    \includegraphics[width=0.8\linewidth]{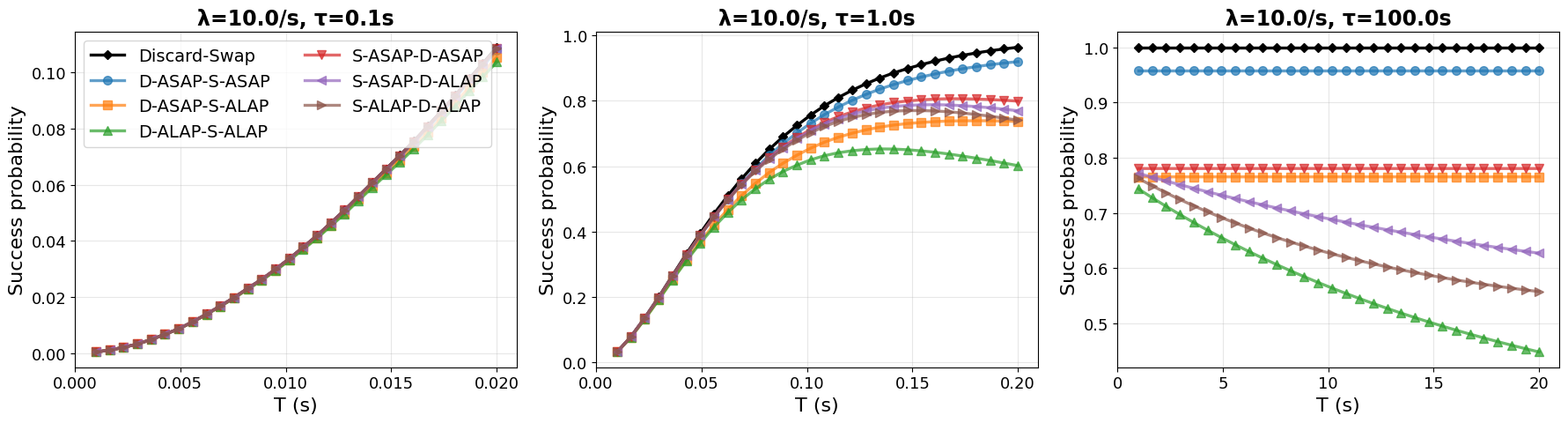}
    \caption{The success probability $\mathrm{Pr}(\text{success})$ for two-hop strategies with $\lambda=10/\mathrm{s}$ and $\tau=0.1, 1, 100\,\mathrm{s}$.}
    \label{fig:double-link-psucc-lam10}
\end{figure*}

\begin{figure*}
    \centering
    \includegraphics[width=0.8\linewidth]{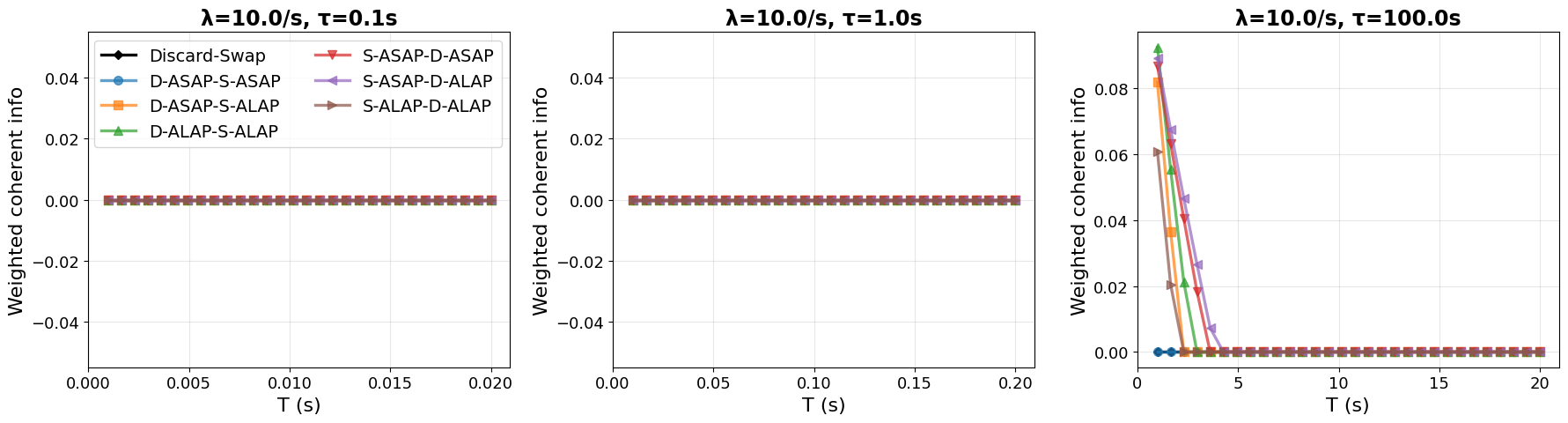}
    \caption{The weighted coherent information $R$ for two-hop strategies with $\lambda=10/\mathrm{s}$ and $\tau=0.1, 1, 100\,\mathrm{s}$.}
    \label{fig:double-link-Ic-lam10}
\end{figure*}

We plot the expected output fidelities achieved by each strategy in Fig.~\ref{fig:double-link-fidelity-lam10}. For short coherence times ($\tau=0.1\ \mathrm{s}$) the seven strategies fall into two nearly degenerate groups: \textsf{S-ASAP-D-ASAP} and \textsf{S-ASAP-D-ALAP} track one another closely at the top, with \textsf{D-ASAP-S-ASAP} either close below or close above the previous two, while \textsf{Discard-Swap}, \textsf{D-ASAP-S-ALAP}, \textsf{D-ALAP-S-ALAP} and \textsf{S-ALAP-D-ALAP} are essentially indistinguishable below them. This is consistent with the one-hop behavior, where the output fidelities of the three single-segment strategies nearly coincide for $T$ that is not too large, so the fidelities of the states immediately before an end-of-window swap are approximately equal. For larger coherence time the strategies separate clearly. In general, swapping ASAP achieves higher expected fidelity than swapping ALAP, because leaving two states idle in noisy memories exposes both to noise, whereas after swapping only one state is subject to noise. Note that the highest expected fidelity and the highest weighted coherent information are not achieved by the same strategy, since $\bar F$ takes no account of how often the strategy succeeds.

The success probability of each strategy is plotted in Fig.~\ref{fig:double-link-psucc-lam10}. \textsf{Discard-Swap} naturally achieves the highest success probability, approaching unity at large $T$, followed by \textsf{D-ASAP-S-ASAP}, which settles near $0.96$ in the high-coherence regime. This is because the causal \textsf{D-ASAP-S-ASAP} policy commits to its swap as soon as both segments are ready and distills only when a second pair happens to be present already, so it attempts fewer distillations than the other strategies and forfeits fewer realizations to distillation failure. For large $T$, the success probability of \textsf{S-ASAP-D-ASAP} approaches a constant that can be approximated by
\begin{align}
    P_\text{dist}\Big(F_\text{swap}\big(F(0),F(1/\lambda)\big),F_\text{swap}\big(F(0),F(1/\lambda)\big)\Big).
\end{align}
The remaining strategies, namely \textsf{S-ASAP-D-ALAP}, \textsf{S-ALAP-D-ALAP}, and \textsf{D-ALAP-S-ALAP}, succeed with probabilities that decrease with time in the high-coherence regime, since the states they must hold until $T$ decohere and the final distillation becomes correspondingly less likely to succeed.

The weighted coherent information of the expected output is plotted in Fig.~\ref{fig:double-link-Ic-lam10}. For small coherence time ($\tau=0.1\ \mathrm{s}$ and $\tau=1\ \mathrm{s}$ at $\lambda=10/\mathrm{s}$) it is identically zero for all strategies, because the achieved fidelities are all below the positive coherent information threshold $F^*\approx0.811$. \textsf{Discard-Swap} achieves zero weighted coherent information in every regime we tested. In fact, the largest fidelity that can possibly be achieved by \textsf{Discard-Swap} even without the effect of noise is $F_\text{swap}(F_0,F_0)=F_0^2+(1-F_0)^2/3\approx0.813$, barely above the threshold $F^*$. \textsf{D-ASAP-S-ASAP} likewise never reaches positive weighted coherent information: its conditional output fidelity settles just below $F^*$, at $\bar F\approx0.809$ at $\lambda=10/\mathrm{s},\ \tau=100\ \mathrm{s},\ T=1\ \mathrm{s}$, because the causal policy most often swaps two undistilled elementary pairs at the first moment both segments are ready.

Among the remaining strategies, the ordering by peak weighted coherent information at $\lambda=10/\mathrm{s},\ \tau=100\ \mathrm{s}$ is \textsf{D-ALAP-S-ALAP} ($R\approx0.092$), \textsf{S-ASAP-D-ALAP} ($0.089$), \textsf{S-ASAP-D-ASAP} ($0.087$), \textsf{D-ASAP-S-ALAP} ($0.082$) and \textsf{S-ALAP-D-ALAP} ($0.061$). The leading strategy is not the same at every operating point. \textsf{D-ALAP-S-ALAP} attains the largest single value, at the shortest deadline plotted, but its weighted coherent information falls off quickly as $T$ grows, and for every larger deadline in this regime \textsf{S-ASAP-D-ALAP} is ahead. The same pattern holds at $\lambda=3/\mathrm{s},\ \tau=100\ \mathrm{s}$, while at $\lambda=100/\mathrm{s},\ \tau=100\ \mathrm{s}$ \textsf{S-ASAP-D-ALAP} leads at every deadline. What is robust across all regimes tested is the weaker statement that the best-performing strategies are those that defer the final distillation to the end of the time window: distillation at $T$ acts on the states in the form in which they will actually be consumed, and avoids paying decoherence on a distilled state that must then idle. Which of \textsf{D-ALAP-S-ALAP} and \textsf{S-ASAP-D-ALAP} is preferable depends on the deadline, and the precise ranking should be checked at any given operating point rather than assumed.

\subsection{Fidelity Distributions}
We again obtain the full fidelity and coherent information distributions from $10^5$ attempted Monte Carlo realizations at fixed parameters ($\lambda=1.0/\mathrm{s}$, $\tau=100.0\,\mathrm{s}$, $T=0.5\,\mathrm{s}$) for representative two-hop strategies.

\begin{figure}
    \centering
    \subfloat[]{\includegraphics[width=0.5\linewidth]{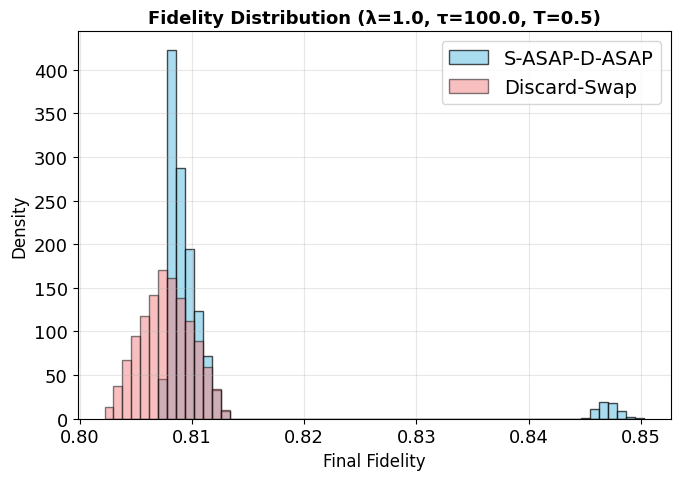} 
    \label{fig:fidelity_histogram}
    }
    \subfloat[]{\includegraphics[width=0.5\linewidth]{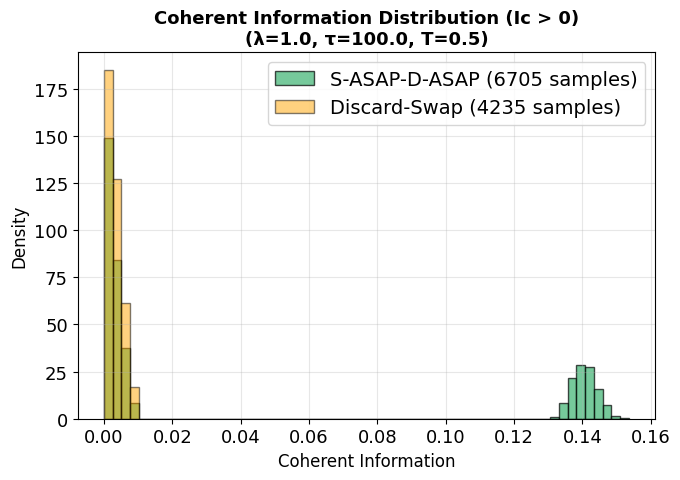}
    \label{fig:coherent_info_histogram}
    }
    \caption{(a) Distribution of final swapped fidelities for \textsf{S-ASAP-D-ASAP} and \textsf{Discard-Swap} at fixed parameters ($\lambda=1.0\ \mathrm{s}^{-1}$, $\tau=100.0\ \mathrm{s}$, $T=0.5\ \mathrm{s}$, $10^5$ attempted samples). The \textsf{S-ASAP-D-ASAP} distribution is bimodal, with a narrow principal peak near $F\approx0.808$ and a small isolated cluster near $F\approx0.847$ from the realizations in which a second full swap and the subsequent distillation both succeeded. \textsf{Discard-Swap} is unimodal and broader, centered near $F\approx0.807$, with no counterpart to the upper cluster. Both data sets share common bin edges. (b) Distribution of coherent information conditioned on $I_c > 0$ at the same parameters. Of $10^5$ attempted samples, $6{,}705$ (\textsf{S-ASAP-D-ASAP}) and $4{,}235$ (\textsf{Discard-Swap}) successful outputs have positive $I_c$. The two clusters visible here correspond to the two components of panel (a).}
\end{figure}

Figure~\ref{fig:fidelity_histogram} shows the fidelity distribution for the \textsf{S-ASAP-D-ASAP} strategy. It is bimodal, with a narrow principal peak near $F\approx0.808$ from realizations in which no second full swap was possible, so that the first swapped pair is delivered undistilled, and a small isolated cluster near $F\approx0.847$ from realizations in which a second swap and a subsequent distillation both succeeded. Unlike the one-hop case, distillation here requires a second full swap, that is, a second elementary pair arriving on \emph{both} segments before $T$, since CNOT distillation acts on two copies of an end-to-end entangled state shared by the same two parties; this joint requirement limits how often the higher-fidelity outcome is reached. The \textsf{Discard-Swap} distribution, by contrast, is unimodal and broader, centered near $F\approx0.807$, since it never performs distillation and its output fidelity varies smoothly with the arrival times of the pairs used in swapping.

Figure~\ref{fig:coherent_info_histogram} shows the coherent information distribution conditioned on $I_c > 0$. The threshold $F^* \approx 0.811$ is cleared by $6{,}705$ of $10^5$ realizations for \textsf{S-ASAP-D-ASAP} and $4{,}235$ for \textsf{Discard-Swap}. The conditional distributions inherit the two-component structure of the fidelity distribution: both strategies place most of their positive-$I_c$ weight just above zero, but only \textsf{S-ASAP-D-ASAP} has a second cluster near $I_c\approx0.14$, produced by the distilled realizations. \textsf{Discard-Swap} has no such component, since it never performs distillation and its output fidelity is bounded by $F_\text{swap}(F_0,F_0)\approx0.813$.

\section{Conclusions and Outlook}\label{sec:conclusions}
In this work, we set out to understand entanglement distillation and swapping in multiplexed quantum repeaters with noisy quantum memories. To this end, we analyzed the most fundamental building blocks for larger repeater networks, i.e., a one-hop link and a two-hop chain, and performed a comprehensive comparison of different possible timings and orderings of distillation and swapping operations. For a one-hop link, we analytically derived the expected fidelity, the success probability, and the weighted coherent information of the expected output of three strategies, namely \textsf{Discard-Oldest}, \textsf{Distill-ASAP}, and \textsf{Distill-ALAP}. We found that for low memory coherence times, \textsf{Discard-Oldest} achieves higher expected output fidelity than the distillation-based strategies once the deadline is long enough for the older pair to have decohered. For higher coherence times, distillation-based strategies begin to show advantages, surpassing the fidelity of the original entangled states, with \textsf{Distill-ALAP} achieving the highest expected output fidelity and, over most of the deadline range, the highest weighted coherent information, at the expense of a lower success probability. For the two-hop chain, we performed a Monte Carlo simulation for these quantities, and we found that, assuming entanglement swapping succeeds deterministically, the strategies that defer the final distillation to the end of the time window achieve the highest weighted coherent information, with \textsf{Distill-ALAP-then-Swap-ALAP} attaining the largest single value at short deadlines and \textsf{Swap-ASAP-then-Distill-ALAP} leading at longer ones. Neither \textsf{Discard-Oldest-then-Swap} nor \textsf{Distill-ASAP-then-Swap-ASAP} reaches positive weighted coherent information in any regime we tested. These results provide important operational insights into quantum repeaters that are limited by finite memory times.

We have adopted a few simplifying assumptions to shed light on the fundamental effects of noisy memories on entanglement distillation and swapping. For instance, we have ignored the classical communication time that is inevitable in performing distillation and swapping, and idealized the heralded entanglement generation process by ignoring synchronization constraints. For a more practical analysis of real-world quantum networks, these timing effects should be included. It will be interesting to perform simulations that include these practical considerations using more sophisticated quantum network simulators~\cite{wu2021sequence,coopmans2021netsquid,satoh2022quisp,kimlee2025quantumsavory}. We have also assumed that at most two entangled states can be stored on each link, because the strategy and state spaces of entanglement distillation with more than two entangled states will become much more complex. Given more than two entangled states, it remains unclear what the optimal way to pair them up and distill is, or whether a collective strategy can perform better than pairwise distillation strategies. To this end, new insights into the fundamentals and practicalities of entanglement distillation are required. Moreover, our error model focuses on depolarization in the raw entangled states and quantum memories. We leave the inclusion of gate and measurement errors, and the consideration of biased or even non-Pauli errors for future work. Finally, a natural next step is to combine existing ideas in memory cutoff strategies with the results on entanglement distillation in this work to optimally combat memory decoherence.

\section*{Acknowledgment}
We thank Joaquin Chung and Eric Chitambar for valuable discussions. X.C. is supported by the U.S. Department of Energy Office of Science National Quantum Information Science Research Centers.

\appendices

\section{Analytical Expressions}\label{sect:analytical-expressions}
For completeness, we present the analytical expressions for the one-hop strategies from the integrals given in Sec.~\ref{sec:one-hop}. The success probability of \textsf{Discard-Oldest} is
\begin{align}
    P_\textsf{Disc} = 1-e^{-2\lambda T}.
\end{align}
The overall success probability of \textsf{Distill-ASAP} is given by $P_\textsf{D-ASAP}=P_1+P_{2,\textsf{D-ASAP}}$, where 
\begin{align}
    P_1 = 2e^{-2\lambda T}(e^{\lambda T}-1),
\end{align}
and 
\begin{align}
    P_{2,\textsf{D-ASAP}} = 
    &\frac{e^{-2\left(\frac{1}{\tau} + 2\lambda\right)T}}{9\left(\lambda^2 \tau^2 - 4\right)}
    \Bigg(
    - e^{3\lambda T} (1 - 4F_0)^2 \lambda^2 \tau^2\notag\\
    &+ e^{2\left(\frac{1}{\tau} + \lambda\right)T}
    (2 + \lambda\tau)
    \big((5 - 4F_0 + 8F_0^2)\lambda\tau - 9\big) \notag\\
    &+ e^{2\left(\frac{1}{\tau} + 2\lambda\right)T}
    (\lambda\tau - 2)
    \big(9 + (5 - 4F_0 + 8F_0^2)\lambda\tau \big) \notag\\
    &- 9 e^{\left(\frac{2}{\tau} + 3\lambda\right) T}
    \left(\lambda^2 \tau^2 - 4\right) 
    \Bigg).
\end{align}
Likewise, the success probability of \textsf{Distill-ALAP} is $P_\textsf{D-ALAP}=P_1+P_{2,\textsf{D-ALAP}}$ with 
\begin{align}
    P_{2,\textsf{D-ALAP}} = 
    &\frac{e^{-\left(\frac{4}{\tau} + 2\lambda \right)T}}{18 (\lambda \tau - 2)^2}
    \Bigg(
    (1 - 4F_0)^2 \lambda^2 \tau^2 e^{2\lambda T} \notag\\
    &- 2 (1 - 4F_0)^2 \lambda^2
    \tau^2 e^{\left(\frac{2}{\tau} + \lambda\right)T}  \notag\\
    &+ 2\Big(\left(8F_0^2 - 4F_0 + 5\right)\lambda^2\tau^2 - 18\lambda\tau + 18\Big)
    e^{\frac{4T}{\tau}} \notag\\
    &+ 9 (\lambda\tau - 2)^2
    e^{\left(\frac{4}{\tau} + 2\lambda\right)T} \notag\\
    &- 18 (\lambda\tau - 2)^2
    e^{\left(\frac{4}{\tau} + \lambda\right)T} \Bigg).
\end{align}

Now, the expected fidelities of these strategies are given by
\begin{align}
    \bar{F}_\mc{S} = \frac{I_1+I_\mc{S}}{P_\mc{S}},
\end{align}
where
\begin{align}
    I_1 =& \frac{e^{-2\lambda T}}{2(\lambda \tau - 2)}
    \Big(2 - 4F_0\lambda\tau + 
    (4F_0 - 1)\lambda\tau e^{(\lambda-2/\tau)T} \notag\\
    &+ (\lambda \tau - 2)\, e^{\lambda T}\Big),
\end{align}
\begin{align}
    I_\textsf{Disc} = 
    &\frac{e^{-2\left(\frac{1}{\tau} + \lambda\right)T}}{4(\lambda\tau - 2)(\lambda\tau - 1)}
    \Bigg(
    (4F_0 - 1)\lambda^2\tau^2 e^{2\lambda T} \notag \\
    &+ e^{2\left(\frac{1}{\tau} + \lambda\right)T}
    (\lambda\tau - 2)(\lambda\tau - 1) \notag\\
    &- 4 e^{\left(\frac{2}{\tau} + \lambda\right)T}
    (\lambda\tau - 1)(2F_0\lambda \tau - 1) \notag\\
    &+ e^{\frac{2T}{\tau}}
    (\lambda\tau - 2)(4F_0\lambda \tau - 1)
    \Bigg),
\end{align}
\begin{align}
    I_\textsf{D-ASAP} = 
    &\frac{e^{-2\left(\frac{1}{\tau} + \lambda\right)T}}{36(\lambda\tau - 2)(\lambda\tau - 1)(2 + \lambda\tau)} \notag \\
    & \times \Bigg(
    e^{2\left(\frac{1}{\tau} + \lambda\right)T}
    (\lambda\tau - 2)(\lambda\tau - 1)\nonumber\\
    &~~~~~~~~ \times\big(9 + (5 - 4F_0 + 8F_0^2)\lambda\tau\big) \notag\\
    &~~~~ + e^{2\lambda T} (4F_0 - 1)(2 + \lambda\tau) \notag \\
    &~~~~~~~~ \times \big(1 + \lambda\tau + 8F_0(\lambda\tau - 2)\big) \lambda \tau \notag\\
    &~~~~ - 2e^{\lambda T} (4F_0 - 1)(\lambda\tau - 1) \notag \\
    &~~~~~~~~ \times \big(2F_0(8 + 5\lambda\tau) - \lambda\tau - 1 \big) \lambda \tau \notag\\
    &~~~~ - 6 e^{\left(\frac{2}{\tau} + \lambda\right)T}
    (\lambda\tau - 1)(2 + \lambda\tau) \notag \\
    &~~~~~~~~ \times \big((1 + 2F_0)\lambda\tau - 3\big) \notag\\
    &~~~~ + \big(8F_0(5F_0 - 1) + 4\big) (2 + \lambda\tau) \lambda^2\tau^2 e^{\frac{2T}{\tau}} \notag \\
    &~~~~ - \big(8F_0(F_0 + 1) + 11 \big) (2 + \lambda\tau) \lambda\tau e^{\frac{2T}{\tau}} \notag \\
    &~~~~ + 9 (2 + \lambda\tau)e^{\frac{2T}{\tau}}
    \Bigg),
\end{align}
and
\begin{align}
    I_\textsf{D-ALAP} = 
    &\frac{e^{-2\left(\lambda+2/\tau\right)T}}{72 (\lambda\tau - 2)^2} \Bigg(
    5 (1 - 4F_0)^2 \lambda^2 \tau^2 e^{2\lambda T} \notag \\
    &+ 6 (4F_0 - 1)(\lambda\tau-2)\lambda \tau e^{2T(\lambda+1/\tau)} \notag\\
    &- 4 (4F_0 - 1)\big((10F_0 - 1)\lambda^2\tau^2 - 3\lambda\tau\big)e^{(\lambda+2/\tau)T} \notag\\
    &- 12 (\lambda\tau-2)\big((2F_0 + 1)\lambda\tau - 3\big)
    e^{(\lambda+4/\tau)T} \notag\\
    &+ 9(\lambda\tau-2)^2e^{2(\lambda+2/\tau)T} \notag\\
    &+ 4 \Big(2\lambda\tau \big(2F_0((5F_0 - 1)\lambda\tau - 3) + \lambda\tau - 3\big) + 9\Big)\notag\\
    &~~~~\times e^{4T/\tau} 
    \Bigg).
\end{align}

\bibliography{refs}
\bibliographystyle{IEEEtran}

\end{document}